\documentclass[10pt,english,twocolumn,pra,aps]{revtex4}
\usepackage[T1]{fontenc}
\usepackage[latin1]{inputenc}
\usepackage{amsmath}
\usepackage{graphicx}
\usepackage{amssymb}

\usepackage{amsmath} \usepackage{amsfonts} \usepackage{amssymb}
\usepackage{graphicx} \usepackage[dvips]{hyperref} \usepackage{color}

\makeatletter

\usepackage{amsfonts}

\usepackage[dvips]{hyperref}

\usepackage{babel}
\makeatother

\definecolor{plum}{rgb}{0.5,0,1} \definecolor{darkgreen}{rgb}{0,0.5,0}
\definecolor{pink}{rgb}{1,0,.5} \definecolor{orange}{rgb}{.5,1,0}
\newcommand{\killed}[1]{} 

\begin{document}

\title{Cold Atoms and Molecules in Self-Assembled Dipolar Lattices}

\date{\today}

\author{G. Pupillo$^{1,2}$, A. Griessner$^{1,2}$, A. Micheli$^{1,2}$,
M. Ortner$^{1,2}$, D.-W. Wang$^{3}$ and P. Zoller$^{1,2}$}

\affiliation{$^{1}$Institute for Theoretical Physics, University of Innsbruck,
A-6020, Innsbruck, Austria\\
 $^{2}$Institute for Quantum Optics and Quantum Information of the
Austrian Academy of Sciences, A-6020, Innsbruck, Austria\\
 $^{3}$Physics Department and NCTS, National Tsing-Hua University, Hinschu, Taiwan, ROC}

\begin{abstract}
We study the realization of lattice models, where cold atoms and
molecules move as extra particles in a dipolar crystal of trapped
polar molecules. The crystal is a self-assembled floating mesoscopic
lattice structure with quantum dynamics given by phonons. We show
that within an experimentally accessible parameter regime extended
Hubbard models with tunable long-range phonon-mediated interactions
describe the effective dynamics of dressed particles.
\end{abstract}
\maketitle Trapped atomic and molecular quantum gases allow the
realization of quantum lattice models of strongly interacting
bosonic and fermionic particles. For example, the dynamics of atoms
in optical lattices is well described by a Hubbard model, where the
tunability of the Hubbard parameters via external fields combined
with atomic physics techniques of preparation and measurement
provides a \emph{quantum simulator} of strongly correlated condensed
matter models \cite{0}. In this letter we propose and study an
alternative scenario of realizing lattice models, where a dipolar
crystal of trapped polar molecules provides a self-assembled
floating lattice structure for extra particles, which are atoms or
molecules of a second species (Fig. 1). By confining polar molecules
to an effective 2D (Fig. 1a) or 1D (Fig.1b,c) geometry by strong
transverse trapping, dipolar crystals can form as a result of the
balance of strong repulsion between the dipoles aligned by an
external electric field, and an in-plane trapping potential
\cite{1,Exp}. Particles moving in this lattice under conditions of
elastic scattering see a periodic potential, and thus form a lattice
gas.

The distinguishing features of this realization of lattice models
are: (i) Dipolar molecular crystals constitute an array of
microtraps with its own quantum dynamics represented by phonons
(lattice vibrations), while the lattice spacings are tunable with
external control fields, ranging from a $\mu$m down to the hundred
nm regime, i.e. potentially smaller than for optical lattices. (ii)
The motion of the extra particles is governed by an interplay of
\emph{Hubbard (correlation) dynamics} in the lattice and
\emph{coupling to phonons}. The tunability of the lattice allows to
access a wide range of Hubbard parameters and phonon couplings.
Compared with optical lattices, for example, a small scale lattice
yields significantly enhanced hopping amplitudes, which set the
relevant energy scale for our Hubbard model (e.g. for exchange
interactions), and thus also the temperature requirements for
realizing strongly correlated quantum phases. While the molecular
setups we propose are reminiscent of (and may be relevant to) solid
state systems with strong phonon couplings, as e.g. polaronic and/or
superconducting materials, they realize the unusual parameter regime
where the mass of the crystal's and of extra particles can be
comparable.
%
%
\begin{figure}[t!]
\begin{centering}
\includegraphics[width=1\columnwidth]{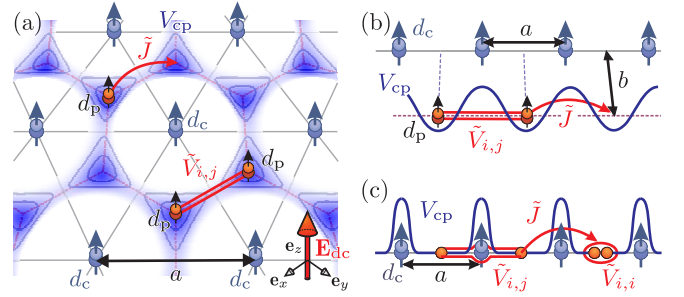}
\par\end{centering}
\caption{\label{figs:fig1}A dipolar crystal of polar molecules in 2D
(a) and 1D (b,c) provides a periodic lattice $V_{{\rm cp}}$ for
extra atoms or molecules giving rise to a lattice model with hopping
$\tilde{J}$ and long-range interactions $\tilde{V}_{i,j}$ (see
text). (a) In 2D a triangular lattice is formed by polar molecules
with dipole moment $d_{{\rm c}}$ perpendicular to the plane. A
second molecular species with dipole moment $d_{{\rm p}}\ll d_{{\rm
c}}$ moves in the honeycomb lattice $V_{{\rm cp}}$ (darker shading
corresponds to deeper potentials). (b) A 1D dipolar crystal with
lattice spacing $a$ provides a periodic potential for a second
molecular species moving in a parallel tube at distance $b$ (Setup
1). (c) 1D setup with atoms scattering from the dipolar lattice
(\textit{\emph{Setup 2}}).}
\end{figure}

A homogeneous lattice of polar molecules underlying the
configurations of Fig. 1 relies on the strong repulsive
dipole-dipole interactions $V_{{\rm c}}({\bf R})=d_{{\rm
c}}^{2}/R^{3}$ with $R$ the distance between the molecules, and
$d_{\rm c}$ the dipole moment induced by a transverse electric field
$E_{{\rm dc}}$. A requirement for the existence of a crystal is that
the ratio of the potential energy to the kinetic energy of small
oscillations around the equilibrium position, $r_{d}=d_{{\rm
c}}^{2}m_{{\rm c}}/\hbar^{2}a$, with $a$ the lattice spacing and
$m_{{\rm c}}$ the mass, is larger than a critical value $r_{c}$,
where $r_{c}=18\pm4$ and $r_{c}\sim1$ for bosons at zero temperature
in 2D and 1D, respectively \cite{1}. Thus a dipolar crystal will
form for $a<a_{\textrm{max}}\equiv d_{{\rm
c}}^{2}m_{c}/\hbar^{2}r_{c}$. In addition, we have
$a_{\textrm{min}}<a$ with $a_{\textrm{min}}=(12d_{{\rm
c}}^{2}/m_{{\rm c}}\omega_{\perp}^{2})^{1/5}$, which reflects the
requirement of strong transverse trapping with a harmonic oscillator
frequency $\omega_{\perp}$ to prevent collapse due to attractive
forces between aligned dipoles. For RbCs (SrO) molecules with
permanent dipole moment $d_{\rm c}=1.27$D ($d_{{\rm c}}=8.9$D)
confined by an optical lattice with $\omega_{\perp}/2\pi\sim150$kHz,
$a_{\textrm{min}}\sim100$nm ($200$nm), while $a_{\textrm{max}}$ can
be several $\mu$m. Excitations of the crystal are acoustic phonons
with Hamiltonian $H_{{\rm c}}=\sum_{q}\hbar\omega_{ q}a_{
q}^{\dagger}a_{q}$, where $a_{q}$ destroys a phonon of quasimomentum
${\bf q}$ in the mode $\lambda$. In 1D,
$\hbar\omega_{q}=(2/\pi^{2})\left[12r_{d}f_{q}\right]^{1/2}E_{{\rm
R,c}}$ with Debye frequency $\hbar\omega_{{\rm
D}}\equiv\hbar\omega_{\pi/a}\sim1.4\sqrt{r_{d}}E_{{\rm R,c}}$,
$f_{q}=\sum_{j>0}4\sin(qaj/2)^{2}/j^{5}$, and lattice recoil
frequency $E_{{\rm R,c}}\equiv\hbar^{2}\pi^{2}/2m_{{\rm c}}a^{2}$
(typically a few to tens of kHz). The classical melting of the
crystal into a normal phase occurs at $k_B T_C\simeq 0.018 r_d
E_{{\rm R,c}}$ and $0.2r_d E_{{\rm R,c}}$ in 2D and 1D, respectively
\cite{1,Kalia}.

An extra particle confined to the 2D crystal plane (Fig. 1a) or a 1D
tube (Fig. 1b,c) will scatter from the periodic lattice potential
$\sum_{j}V_{{\rm cp}}(\mathbf{R}_{j}-\mathbf{r})$ with $\mathbf{r}$
and $\mathbf{R}_{j}$ the coordinates of the particle and crystal
molecule $j$, respectively. We write ${\bf \mathbf{R}}_{j}^{}={\bf
\mathbf{R}}_{j}^{0}+{\bf \mathbf{u}}_{j}$ with $\mathbf{R}_{j}^{0}$
the equilibrium positions and $\mathbf{u}_{j}$ small displacements,
assuming that the particles do not significantly perturb the
lattice. For particles being molecules, this potential is given by
the repulsive dipole-dipole interaction $V_{{\rm cp}}({\bf
R}_{j}-{\bf r})=d_{{\rm p}}d_{{\rm
c}}/\mid\mathbf{R}_{j}-\mathbf{r}\mid^{3}$ with $d_{{\rm p}}\ll
d_{{\rm c}}$ the induced dipole moment, and in the case of atoms we
assume that the interaction can be modeled by a short range
pseudopotential proportional to an elastic scattering length
$a_{{\rm cp}}$. In addition, extra molecules and atoms will interact
according to dipolar, or short range interactions, respectively.

We consider a situation where the dynamics of the extra particles in
the lattice can be described by a single band Hubbard Hamiltonian
coupled to the acoustic phonons of the lattice \cite{Mahan} 

\begin{eqnarray}
H & = & -J\sum_{<i,j>}c_{i}^{\dagger}c_{j}
+\tfrac{1}{2}\sum_{i,j}V_{ij}c_{i}^{\dagger}c_{j}^{\dagger}c_{j}c_{i}\nonumber \\
 & + & \sum_{q,j}M_{ q}e^{i{\bf q}\cdot{\bf R}_{j}^{0}}c_{j}^{\dagger}c_{j}(a_{q}
 +a_{-q}^{\dagger})+H_{{\rm c}}.\label{eq:eqSmallPolaron}\end{eqnarray}
 The first line describes the nearest neighbor hopping of the extra
 particles
with hopping amplitudes $J$, and interactions $V$. We denote by
$c_{i}$ ($c_{i}^{\dagger}$) destruction (creation) operators of the
particles. The first term in the second line is the phonon coupling
obtained in lowest order in the displacement ${\bf
u}_{j}=i\sum_{q}(\hbar/2m_{{\rm c}}N\omega_{ q})^{1/2}\xi_{
q}(a_{q}+a_{- q}^{\dagger})e^{i{\bf q}\cdot{\bf R}_{j}^{0}}$ with
$M_{q}=\bar{V}_{{\bf q}}{\bf q}\cdot\xi_{ q}(\hbar/2Nm_{{\rm
c}}\omega_{ q})^{1/2}\beta_{{\bf q}}$, where $\xi_{ q}$ and $N$ are
the phonon polarization and the number of lattice molecules,
respectively, $\bar{V}_{{\bf q}}$ is the Fourier transform of the
particle-crystal interaction $V_{\rm cp}$, and $\beta_{{\bf q}}=\int
d\mathbf{r}|w_{0}({\bf \mathbf{r}})|^{2}e^{i\mathbf{qr}}$, with
$w_{0}({\bf \mathbf{r}})$ the Wannier function of the lowest Bloch
band \cite{Mahan}. The validity of the single band Hubbard model
requires $J,V<\Delta$, and temperatures $k_{B}T<\Delta$ with
$\Delta$ the separation to the first excited Bloch band. We note
that the Hubbard parameters are of the order of magnitude of the
recoil energy, $J,V\sim E_{{\rm R,c}}$, and thus they are (much)
smaller than the Debye frequency $\hbar\omega_{D}\sim E_{{\rm
R,c}}\sqrt{r_{d}}$, for $r_d\gg1$ \cite{Albus}. 

Below we will present detailed results for the examples of
Figs.~1b,c. The separation of time scales $J,V$$\ll\hbar\omega_{D}$,
combined with the fact that the coupling to phonons is dominated by
high frequencies $\hbar\omega>J,V$ (see the discussion of $M_{ q}$
below) is reminiscent of polarons as particles dressed by (optical)
phonons, where the dynamics is given by coherent and incoherent
hopping on a lattice \cite{Mahan,Alexandrov}. This physical picture
is brought out in a master equation treatment within a strong
coupling perturbation theory. The starting point is a Lang-Firsov
transformation of the Hamiltonian $H\rightarrow
\mathcal{S}H\mathcal{S}^{\dagger}$ with a density-dependent
displacement $\mathcal{S}=\exp[-\sum_{ q,j}(M_{ q}/\hbar\omega_{
q})e^{i\mathbf{q}\mathbf{R}_{j}^{0}}c_{j}^{\dagger}c_{j}(a_{ q}-a_{-
q}^{\dagger})]$. This eliminates the phonon coupling in the second
line of Eq.~\eqref{eq:eqSmallPolaron} in favor of a transformed
kinetic energy term $-J\sum_{
<i,j>}c_{i}^{\dagger}c_{j}X_{i}^{\dagger}X_{j}$, where the operators
$X_{j}=\exp[\sum_{ q}M_{ q}e^{i\mathbf{q}\mathbf{R}_{j}^{0}}(a_{
q}-a_{- q}^{\dagger})/\hbar\omega_{ q}]$ can be interpreted as a
lattice recoil of the dressed particles in a hopping process. In
addition, the bare interactions are renormalized according to
$\tilde{V}_{ij}=V_{ij}+V_{ij}^{(1)}$ with $V_{ij}^{(1)}=-
2\sum_{q}\cos({\bf q}({\bf R}_{i}^{0}-{\bf R}_{j}^{0}))M_{
q}^{2}/\hbar\omega_{ q}$, that is, the phonon couplings induce and
modify off-site interactions. The onsite interaction is given by
$\tilde{V}_{j,j}=V_{j,j}-2E_{p}$ with $E_{p}=\sum_{ q}M_{
q}^{2}/\hbar\omega_{ q}$ the \textit{polaron} self-energy or
\textit{polaron shift}. For $J=0$ the new Hamiltonian is diagonal
and describes interacting polarons and independent phonons. The
latter are vibrations of the lattice molecules around new
equilibrium positions with unchanged frequencies. For the models of
Fig.~1 b,c studied below, consistency with the assumption of small
perturbation of the lattice by the extra particles requires $\Delta
u/a= \sum_{q}(8 \hbar^2/m_{\rm c} N )^{1/2}M_{
q}\sin(qa/2)^{2}/(\hbar\omega_{ q})^{3/2}a \ll 1$, which is achieved
for $(E_{p}E_{{\rm R,c}})^{1/2}\ll\hbar\omega_{D}$, and is generally
satisfied.

A Born-Markov approximation with the transformed kinetic energy as
perturbation, and the phonons a finite temperature heatbath with
$J,V\ll \hbar\omega_D$ (see above), provides us with the master
equation for the reduced density operator of the dressed particles
$\rho(t)$ in Lindblad form,\begin{eqnarray}
\dot{\rho}(t) & = & -\frac{i}{\hbar}[\tilde{H}+\sum_{j, l,{\bf \delta}{\bf \delta}'}\Delta_{j l}^{{\bf \delta}{\bf \delta}'}b_{j{\bf \delta}}b_{ l{\bf \delta}'},\rho(t)]\label{lindblad}\\
 &+&\sum_{j,l,{\bf \delta},{\bf \delta}'}\frac{\Gamma_{j, l}^{{\bf \delta},{\bf \delta}'}}{2\hbar}(2b_{j{\bf \delta}}\rho(t)b_{ l{\bf \delta}'}-b_{j{\bf \delta}}b_{l{\bf \delta}'}\rho(t)-\rho(t)b_{j{\bf \delta}}b_{l{\bf \delta}'}),\nonumber \end{eqnarray}
with $b_{j{\bf \delta}}=c_{j+{\bf \delta}}^{\dagger}c_{j}$, and
where the effective system Hamiltonian becomes an extended Hubbard
model,\begin{equation} \tilde{H}=-\tilde{J}\sum_{
<i,j>}c_{i}^{\dagger}c_{j}+\tfrac{1}{2}\sum_{i,j}\tilde{V}_{ij}c_{i}^{\dagger}c_{j}^{\dagger}c_{j}c_{i}.\label{eq:Heff}\end{equation}
\emph{Coherent} hopping of the dressed particles is described by
$\tilde{J}=J\langle\langle X_{i}^{\dagger}X_{j}\rangle\rangle\equiv
J\exp(-S_{T})$, where $S_{T}=\sum_{q}(M_{ q}/\hbar\omega_{
q})^{2}[1-\cos({\bf q}{\bf a})](2n_{ q}(T)+1)]$ characterizes the
strength of the particle-phonon interactions, and $n_{q}(T)$ is the
thermal occupation at temperature $T$ \cite{Mahan}. The terms
involving $\Delta_{j,l}^{{\bf \delta},{\bf \delta}'}$ are second
order corrections, which are small relative to $\tilde{H}$ both in
the \char`\"{}weak\char`\"{} $S_{T=0}\equiv S_{0}\lesssim1$ and
\char`\"{}strong\char`\"{} $S_{0}\gg1$ coupling regimes, provided
$J\ll\hbar\omega_{D}$ and $J\ll E_{p}$, respectively
\cite{Ortner,Alexandrov},  \cite{SecondOrder}. The dissipative term
in Lindblad form in the second line of Eq.~\eqref{lindblad}
corresponds to thermally activated \emph{incoherent} hopping with
rates $\Gamma_{j,l}^{{\bf \delta},{\bf \delta}'}$, which are small
compared to $J$ for $S_T\ll 1$ and  $S_T\gg 1$, provided $J
\bar{V}_{q=0}^2 k_B T/[(\hbar \omega_D)^4 \sqrt{r_d}]\ll 1$ and $k_B
T/E_p\ll 1$, respectively \cite{Ortner,Alexandrov}, and in
particular they are negligible for the energies of interest
$k_{B}T\ll\min(\Delta,E_p,k_B T_{C})$.

In the parameter regime of interest the dynamics of the dressed
particles is described by the extended Hubbard Hamiltonian
$\tilde{H}$. In the following we calculate the effective Hubbard
parameters from the microscopic model for the 1D setups described in
Figs.~1b and c \cite{Mathey04}.
\begin{center}
\begin{figure}[h]
\begin{centering}
\includegraphics[width=1\columnwidth]{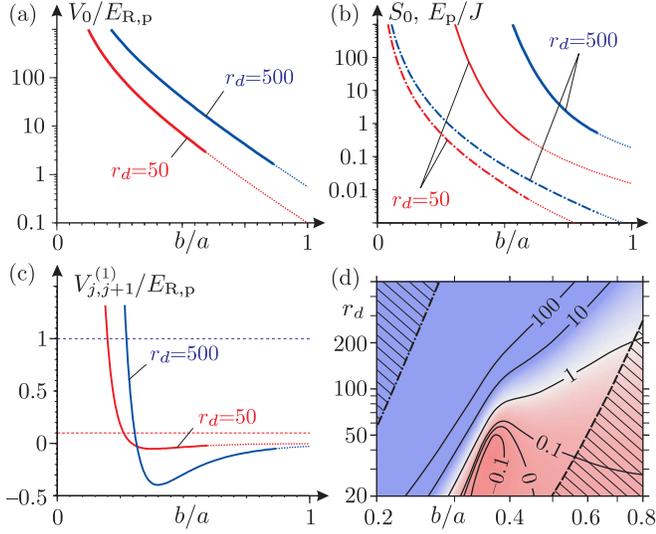}
\par\end{centering}
\caption{\label{figs:fig02} Setup 1 (Fig.~1b): Hubbard parameters
for $d_{{\rm p}}/d_{{\rm c}}=0.1$ and $m_{{\rm c}}=m_{{\rm p}}$. (a)
Lattice depth $V_{0}$ in units of $E_{{\rm R,c}}$ vs.~$b/a$ for
$r_{d}=50$ and $500$. Thick continuous lines: tight-binding regime
$4J<\Delta$. (b) Reduction factor $S_{0}$ (dashed dotted lines) and
polaron shift $E_{p}/J$ (solid lines), for $4J<\Delta$. (c)
Continuous lines: phonon-mediated interactions $V_{j,j+1}^{(1)}$.
Horizontal (dashed) lines: $V_{j,j+1}$. (d) Contour plot of
$\tilde{V}_{j,j+1}/2\tilde{J}$ (solid lines) as a function of $b/a$
and $r_{d}$. A single-band Hubbard model is valid left of the dashed
line ($4J,V_{ij}<\Delta$), and right of the dot-dashed line
($E_{p}<\Delta$).}
\end{figure}
\par\end{center}

\textit{\emph{In}} \textit{Setup 1} (Fig.~1b\textit{\emph{)
molecules}} of a second species are trapped in a tube at a distance
$b$ from the crystal tube under 1D trapping conditions. For crystal
molecules fixed at the equilibrium positions with lattice spacing
$a$, the extra particles feel a periodic potential $V_{{\rm
cp}}(x)=d_{{\rm c}}d_{{\rm
p}}\sum_{j}\left[b^{2}+(x-ja)^{2}\right]^{-3/2}$, which determines
the bandstructure. The potential is sinusoidal for $b/a\gtrsim1/4$,
while for $b/a<1/4$ it has a comb-like structure, since the
particles resolve the individual molecules forming the crystal. The
lattice depth $V_{0}\equiv V_{{\rm cp}}(a/2)-V_{{\rm cp}}(0)$ is
shown in Fig. 2a as a function of $b/a$, where the thick solid lines
indicate the parameter regime $4J<\Delta$. The strong dipole-dipole
repulsion between the extra particles acts as an effective hard-core
constraint \cite{BuchlerNature}. We find that for $4J<\Delta$ and
$d_{{\rm p}}\ll d_{{\rm c}}$ the bare off-site interactions satisfy
$V_{ij}\sim d_{{\rm p}}^{2}/(a|i-j|)^{3}<\Delta$.

The particle-phonon coupling is 
\[ M_{q}=\frac{d_{{\rm c}}d_{{\rm
I}}}{a b}\sqrt{\frac{2\hbar}{N m_{{\rm
c}}\omega_q}}q^2\mathcal{K}_{1}(b|q|)\beta_q\]
 with $\mathcal{K}_{1}$ the modified Bessel function of the second
kind, and $M_{q}\sim\sqrt{q}$ for $q\rightarrow0$. For $b/a<1$,
which is the regime of interest (compare Fig. 2), $M_{q}$ is peaked
at large $q\sim\pi/a$, so that the main contribution to the
integrals in the definition of $S_{T}$ and $E_{p}$ is indeed
dominated by large frequencies $\hbar\omega_{q}>J$. A plot of
$S_{0}$ as a function of $b/a$ is shown in Fig.~2a. We find the
scaling $S_{0}\propto\sqrt{r_{d}}(d_{{\rm p}}/d_{{\rm c}})^{2}$, and
within the regime of validity of the single band approximation,
$S_{0}$ can be tuned from $S_{0}\ll1$ ($\tilde{J}\sim J$) to
$S_{0}\gg1$ ($\tilde{J}\ll J$) corresponding to the large and small
polaron limit, respectively. The polaron shift $E_{p}$, as shown in
Fig. 2b, generally exceeds the bare hopping rate $J$, and in
particular, $E_{p}\gg J$ for $S_{0}\gtrsim1$. Together with the
condition $\hbar\omega_{D}\gg J$ this ensures that the second-order
shifts in the master equation, $\Delta_{i,j}^{\delta,\delta'}$, are
indeed negligible in the parameter regime of interest.

The phonon-mediated interactions $V_{i,j}^{(1)}$ show oscillations,
which for $b/a\lesssim1/4$ decay slowly as  $\sim1/|i-j|^{2}$. These
interactions are thus long-ranged, and, depending on their sign,
they can enhance or reduce the direct dipole-dipole repulsion of the
extra particles. The term $V_{j,j+1}^{(1)}$ is shown in Fig. 2c to
alternate between attractive and repulsive as a function of $b/a$.
The effective Hubbard parameters $\tilde{V}_{j,j+1}$ and $\tilde{J}$
are summarized in Fig.~\ref{figs:fig02}d, which is a contour plot of
$\tilde{V}_{j,j+1}/2\tilde{J}$ as a function of $r_{d}$ and $b/a$.
The ratio $\tilde{V}_{j,j+1}/2\tilde{J}$ increases by decreasing
$b/a$ or increasing $r_{d}$, and can be much larger than one.
Equation~\eqref{eq:Heff} is valid in the region left of the dashed
line, where $4J,V<\Delta$, and right of the dashed-dotted line,
where $E_{p}<\Delta$. For $E_{p}>\Delta$ a multi-band approach is
required.

\textit{\emph{In}} \textit{Setup 2} (Fig.~1c\textit{\emph{) neutral
atoms are trapped in the same tube as}} the crystal molecules. For
fixed molecules providing the lattice structure each atom feels the
1D potential $V_{{\rm cp}}(x)=\sum_{j}g_{{\rm cp}}\delta(x-ja)$,
which determines the bandstructure (Kronig-Penney model). Here
$g_{{\rm cp}}=2\hbar^{2}a_{{\rm cp}}/\mu_{{\rm p}}a_{\perp,{\rm
p}}^{2}$ for a 3D scattering length smaller than the transverse
confinement, $a_{{\rm cp}}\ll a_{\perp, {\rm p}}=(\hbar/m_{\rm p}
\omega_{\perp,{\rm p}})^{1/2}$, with $\mu$ the reduced mass
\cite{Olshanii98}. For $g_{{\rm cp}}/aE_{{\rm R,p}}\gg1$ the width
of the lowest band is $4J\simeq(g_{{\rm cp}}/aE_{{\rm R,p}})^{-1}$,
while the gap $\Delta$ tends to $\Delta\simeq3E_{{\rm R,p}}$, with
$E_{{\rm R,p}}=\hbar^{2}\pi^{2}/2m_{{\rm p}}a^{2}$.

In the following we are interested in bosonic atoms interacting with
each other via a contact potential with coupling strenght $g_{{\rm
pp}}$ determined by their 3D scattering length $a_{{\rm pp}}$, which
is tunable independent of $a_{{\rm cp}}$. The bare Hubbard
interactions are dominated by onsite interactions,
$V_{i,j}\simeq\delta_{ij}U$, where $U=3g_{{\rm pp}}/2a$ for large
$g_{{\rm cp}}$. For the choice of parameters of Fig.~3 a description
in terms of a single-band Hubbard model is valid for $g_{{\rm
cp}}/aE_{{\rm R,p}}\gtrsim0.5$ so that $J,U\ll\Delta$.
\begin{center}
\begin{figure}[t]
\begin{centering}
\includegraphics[width=1\columnwidth]{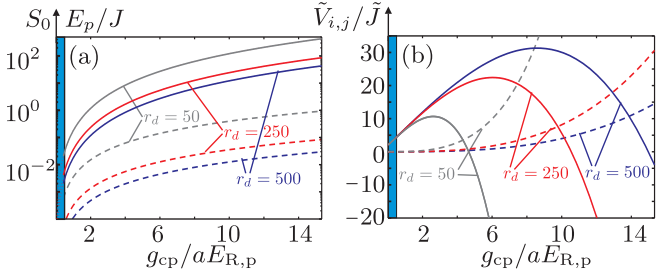}
\par\end{centering}
\caption{Setup 2 (Fig.~\ref{figs:fig1}c): Hubbard parameters for
$m_{{\rm p}}/m_{{\rm c}}=0.814$, $a_{{\rm pp}}=2nm$ and
$\omega_{\perp,{\rm c}}/2\pi=200$kHz. (a) Reduction factor $S_0$
(dashed lines) and polaron shift $E_p/J$ (solid lines) as a function
of the particle-phonon coupling strength $g/a E_{{\rm R},{\rm p}}$.
The tight binding region, $4J, V_{j,j}<\Delta$, is right of the
shaded area. 
(b) Effective onsite ${\tilde V_{j,j}}/{\tilde J}$ (solid lines) and
nearest-neighbor ${\tilde V_{j,j+1}}/{\tilde J} $ (dashed lines)
parameters.}\label{figs:fig03}
\end{figure}
\par\end{center}

The particle-phonon coupling is \[ M_{q}=\frac{g_{{\rm
cp}}}{a}\sqrt{\frac{2\hbar}{Nm_{{\rm c}}\omega_{q}}}|q|\beta_q,\]
which is peaked at large $q\sim\pi/a$, and $M_{q}\sim\sqrt{q}$ for
$q\rightarrow0$. We find the reduction factor $S_{0}\sim
0.92(g_{{\rm cp}}/aE_{{\rm R,p}})^{2}(m_{{\rm c}}/m_{{\rm
p}})^{2}/r_{d}^{3/2}$ and the polaron shift $E_{p}\sim 0.94E_{{\rm
R,p}}(g_{{\rm cp}}/aE_{{\rm R,p}})^{2}(m_{{\rm c}}/m_{{\rm
p}})/r_{d}$, which \emph{decrease} with increasing $r_{d}$ (see
Fig.~3a). For the parameters of Fig.~3 we have $S_{0}\ll1$ over a
wide range of $r_{d}$, so that $\tilde{J}\approx J$. The strong
coupling regime $S_{0}\gg1$, with $E_{p}/J\gg1$, can be reached by
decreasing $r_{d}$ or increasing $g_{{\rm cp}}/aE_{{\rm R,p}}$. We
note that $g_{{\rm cp}}/aE_{{\rm R,p}}$ is restricted by the
condition of a stable crystal $\Delta u\ll a$, i.e. $g_{{\rm
cp}}/aE_{{\rm R,p}}\ll 3r_{d}$.

While the bare atom-atom interaction provides only an onsite shift,
the phonon coupling induces long-range interactions
$V_{i,j}^{(1)},$which decay as $\sim1/|i-j|^{2}$. The effective
interactions $\tilde{V}_{j,j}$, $\tilde{V}_{j,,j+1}$ and $\tilde{J}$
are summarized in Fig.~3b, where the nearest-neighbor term
$\tilde{V}_{j,j+1}\sim0.16E_{p}$ is shown to be repulsive for all
values of $g_{{\rm cp}}/aE_{{\rm R,p}}$, while the onsite
interaction $\tilde{V}_{j,j}=U-2E_{p}$ turns from positive to
negative, which for bosons indicates an instability towards
collapse. That is, for stability we require $g_{{\rm pp}}/aE_{{\rm
R,p}}\gtrsim1.4(g_{{\rm cp}}/aE_{{\rm R,p}})^{2}(m_{{\rm c}}/m_{{\rm
p}})^{2}/r_{d}$. Fig.~3b shows that a regime of strong interactions
$\tilde{V}_{j,j},\tilde{V}_{j,j+1}\gg\tilde{J}$ can be reached for
$r_{d}\gg1$.

One feature of our extended Hubbard model is the appearance and
tunability of strong off-site interactions, a necessary ingredient
for a variety of new quantum phases
\cite{Goral02,Baranov05}, which is difficult to realize in a
standard atomic setup \cite{Bruderer07}. As an example, at half
filling the particles in Setup 1 undergo a transition from a
(Luttinger) liquid ($\tilde{V}_{i,i+1}<2\tilde{J}$) to a
charge-density-wave (CDW) ($\tilde{V}_{i,i+1}>2\tilde{J}$), which
can be observed, e.g. for $r_{d}=100$ at $b/a\approx0.5$ (see Fig.
2d) \cite{Hirsch82}. Similarly, the groundstate of Setup 2 at half
filling is a CDW for $\tilde{V}_{j,j},\tilde{V}_{j,j+1}>4\tilde{J}$,
\cite{Hirsch82}. For larger filling and strong interactions
$2\tilde{V}_{j,j+1}\gtrsim\tilde{V}_{j,j}>4\tilde{J}$, the system
undergoes a second order transition to a supersolid phase, where
diagonal and off-diagonal orders coexist. These strong interactions
are here realized, e.g., for $r_{d}=250$ at $g_{{\rm cp}}/aE_{{\rm
R,p}}\approx 10$, where $\tilde{V}_{j,j+1}\approx V_{j,j}\approx
7.6\tilde{J}$ (see Fig.~3b).

In conclusion, we have studied a scenario where cold atoms or
molecules move in the periodic potential provided by a dipolar
molecular crystal, with quantum dynamics given by phonons. This is
in contrast to familiar traps in atomic physics, where backaction is
negligible. Strong phonon-mediated off-site interactions and
particle localization open new perspectives for studying the
interplay between strong correlations and phonon dynamics in a
tunable setup.

The authors thank H.P.~Büchler and K.~Hammerer for discussions. This
work was supported by the Austrian Science Foundation, the EU under
grants FP6-013501-OLAQUI, MRTN-CT-2003-505089, SCALA IST-15714, and
the Institute for Quantum Information. DWW acknowledges the support
of NSC through NCTS.


\begin{thebibliography}{10}

\bibitem{0}{See e.g.: M. Lewenstein \textit{et al.}, Adv. Phys. {\bf 56}, 243 (2007),
and references therein; D. Jaksch and P. Zoller, Ann. Phys. {\bf
315}, 52 (2005).}

\bibitem{1}{H.P.~B\"uchler~\textit{et al.}, Phys. Rev. Lett. \textbf{98},
060404 (2007); G.E.~Astrakharchik~\textit{et al.}, {\it ibid. }
\textbf{98}, 060405 (2007); A.~S.~Arkhipov~\textit{et al.}, JETP
\textbf{82}, 41 (2005); R.~Citro~\textit{et al.}, Phys. Rev. A {\bf
75}, 051602(R) (2007); A.~Micheli~\textit{et al.}, quant-ph/0703031;
P.~Rabl and P. Zoller, arXiv:0706.3051.}

\bibitem{Exp}{For experiments with polar molecules, see e.g. D.~Wang~\textit{et al.}, Phys. Rev. Lett.
{\bf 93}, 243005 (2004); J.M.~Sage~\textit{et al.}, {\it ibid.} {\bf
94}, 203001 (2005); T.~Rieger~\textit{et al.}, {\it ibid.} {\bf 95},
173002 (2005); S.~Hoekstra~\textit{et al.}, {\it ibid.} {\bf 98},
133001 (2007); W.C.~Campbell~\textit{et al.}, {\it ibid.} {\bf 98},
213001 (2007); B.C.~Sawyer~\textit{et al.}, {\it ibid.} {\bf 98},
253002 (2007).}


\bibitem{Kalia}{R.K.~Kalia and P.~Vashishta, J. Phys. C \textbf{14},
L643 (1981).}

\bibitem{Mahan}{G.D.~Mahan, \textit{Many Particle Physics}, Kluwer
Academic/Plenum Publishers, New York (2000).}

\bibitem{Albus}{This anti-adiabatic regime is hard to achieve in atom-atom mixtures, see e.g.
F.~Illuminati and A.~Albus, Phys. Rev. Lett. \textbf{93}, 090406;
D.-W.~Wang, M.D.~Lukin, and E.~Demler, Phys. Rev. A \textbf{72},
R051604 (2005).}

\bibitem{Alexandrov}{A.S.~Alexandrov, \textit{Theory of Superconductivity},
IoP Publishing, Philadelphia (2003).}

\bibitem{Ortner}{M.~Ortner~\textit{et al.}, in preparation.}

\bibitem{SecondOrder}{For many particles,
$\Delta_{j+1,j}^{-1,1}$ introduces an off-site interaction, which,
for $S_0\gg 1$, can become larger than $\tilde{J}$. However,
$\Delta_{j+1,j}^{-1,1}/V_{j+1,j}^{(1)}\sim (J/E_p)^2$, and thus this
correction is relevant for regimes where $\tilde{V}_{j+1,j}\sim 0$
only. }

\bibitem{Mathey04}{For a discussion of 1D models in atom-atom mixtures
within a Luttinger Liquid formalism, see e.g., L. Mathey~\textit{et
al.}, Phys. Rev. Lett. \textbf{93}, 120404 (2004).}

\bibitem{BuchlerNature}{H.P.~B\"uchler, A.~Micheli and P.~Zoller, Nature Physics
(2007), in press, arXiv:cond-mat/0703688.}

\bibitem{Olshanii98}{$g_{{\rm cp}}$ is tunable via Feshbach resonances in $a_{{\rm cp}}$,
or confinement induced resonances, see M. Olshanii, Phys. Rev. Lett.
\textbf{81}, 938 (1998); E.L.~Bolda, E.~Tiesinga and P.S.~Julienne,
Phys. Rev. A \textbf{66}, 013403 (2002).}

\bibitem{Goral02}{K.~Goral, L.~Santos, and M.~Lewenstein, Phys. Rev. Lett. {\bf
88}, 170406 (2002); R.~Barnett \textit{et al.}, {\it ibid.} {\bf
96}, 190401 (2006); E.G. Dalla Torre, E. Berg, and E. Altman, {\it
ibid.} \textbf{97}, 260401 (2006).}

\bibitem{Baranov05}{M.A.~Baranov {\it et al.}, Phys. Rev. Lett. {\bf 94}, 070404 (2005);
M.A.~Baranov, H.~Fehrmann and M.~Lewenstein, arXiv:cond-mat/0612592;
D.S.~Petrov {\it et al.}, arXiv:0706.2855.}

\bibitem{Bruderer07}{ L.-M.~Duan, E.~Demler, and M.D.~Lukin, Phys. Rev.
Lett. {\bf 91}, 090402 (2003); V.~Scarola and S.~Das Sarma, {\it
ibid.} {\bf 95}, 033003 (2005); M.~Bruderer~\textit{et al.}, Phys.
Rev. A \textbf{76}, 011605(R) (2007).}

\bibitem{Hirsch82}{J.E.~Hirsch and E.~Fradkin, Phys. Rev. B {\bf 27}, 4302 (1983);
P.~Niyaz {\it et al.}, {\it ibid.} {\bf 50}, 362 (1994).}

\bibitem{24}{G.G. Batrouni, F. Hebert, and R.T. Scalettar, Phys.
Rev. Lett. \textbf{97}, 087209 (2006).}
\end{thebibliography}
\end{document}